\documentclass[aps,prb,floatfix,twocolumn,showpacs]{revtex4}
\usepackage{amsmath}
\usepackage{graphicx}
\usepackage{amsfonts}
\usepackage{amssymb}
\begin{document}

\title{Finite frequency noise of a superconductor/ferromagnet quantum point contact}
\author{Audrey Cottet$^{1,2}$, Benoit Dou\c{c}ot$^{2}$, and Wolfgang Belzig$^{3}$}
\begin{abstract}
We have calculated the finite-frequency current noise of a
superconductor-ferromagnet quantum point contact (SF QPC). This signal is
qualitatively affected by the spin-dependence of interfacial phase shifts
(SDIPS) acquired by electrons upon reflection on the QPC. For a weakly
transparent QPC, noise steps appear at frequencies or voltages determined
directly by the SDIPS. These steps can occur at experimentally accessible
temperatures and frequencies. Finite frequency noise is thus a promising tool
to characterize the scattering properties of a SF QPC.
\end{abstract}
\date{\today}
\affiliation{{$^{1}$ Ecole Normale Sup\'{e}rieure, Laboratoire Pierre Aigrain, 24 rue
Lhomond, 75231 Paris Cedex 05, France}}
\affiliation{{$^{2}$ Laboratoire de Physique Th\'{e}orique et Hautes \'{E}nergies,
Universit\'{e}s Paris 6 et 7, CNRS, UMR 7589, 4 place Jussieu, F-75252 Paris
Cedex 05, France}}
\affiliation{{$^{3}$ Fachbereich Physik, Universit\"at Konstanz, D-78457 Konstanz, Germany}}
\maketitle

Spin-dependent transport in mesoscopic structures with ferromagnetic elements
is generating a growing interest, due to the great variety of behaviors
allowed by the lifting of spin-degeneracy and the possibilities of
applications in the field of nanospintronics\cite{Prinz}. In this context, the
study of finite-frequency noise ($\mathcal{FFN}$) has raised little attention
so far\cite{FF}. However, a circuit is characterized by a larger number of
parameters when spin-degeneracy is lifted, and current noise can thus be
particularly useful to obtain more information on those parameters. Recently,
superconductor/ferromagnet ($SF)$ quantum points contacts (QPCs) have raised
much interest as a possible source of information on the properties of $F$
materials. The average current flowing through $SF$ QPCs has been intensively
studied (see e.g. Ref. \onlinecite{Soulen}), and the zero-frequency current
noise of a single-channel $SF$ QPC has been recently calculated\cite{zfQPC}.
It has been shown that these quantities are affected by the spin-polarization
of the QPC transmission probabilities, but also by the Spin-Dependence of
Interfacial Phase Shifts (SDIPS) acquired by electrons upon reflection on the
$SF$ interface. Regarding $\mathcal{FFN}$ in QPCs, the normal metal/normal
metal ($NN$) case has been studied theoretically\cite{Blanter} and
experimentally\cite{Christian}, and the $SN$ case has been addressed
theoretically\cite{Torres}, but spin-dependent transport has not been
considered so far.

In this Letter, we investigate theoretically the $\mathcal{FFN}$ of a $SF$
QPC. Our main result is that, for a weakly transparent contact, the
$\mathcal{FFN}$ is qualitatively affected by the SDIPS. More precisely, some
characteristic noise steps appear at various frequencies or voltages
determined directly by the SDIPS value. These steps occur at experimentally
accessible temperatures and frequencies. Therefore, $\mathcal{FFN}$ is a
promising tool to characterize the scattering properties of a $SF$ QPC. More
generally, this work illustrates that $\mathcal{FFN}$ can provide an
interesting insight in spin-dependent transport.

We define the non-symmetrized current noise $S(V,\omega)=\int_{-\infty
}^{+\infty}\langle\Delta\widehat{I}(0)\Delta\widehat{I}(t)\rangle e^{i\omega
t}dt$ of a $SF$ contact, with$\ \Delta\widehat{I}(t)=\widehat{I}(t)-\langle
I\rangle$ and $\widehat{I}$ the current operator of the contact. In general,
$S(V,\omega)\neq S(V,-\omega)$, because $S(V,\omega)$ describes photon
emission (absorption) processes between the system and a detector for
$\omega>0$ ($\omega<0$). The excess noise $S_{ex}(\omega)=$ $S(V,\omega
)-S(V=0,\omega)$ is often the relevant observable in experiments
\cite{Christian,LesovikAndOthers,Deblock,Martin,Onac,Billangeon}.
Interestingly, $S_{ex}$ has already been measured for both signs of $\omega
$\cite{Billangeon}. We first consider a general ballistic $SF$ contact, with
$S$ a conventional superconductor described by the Bogoliubov-De Gennes (BdG)
equations \cite{BdG}. A bias voltage $V$ is applied to the contact, so that
the Fermi level of $S[F]$ can be set to $\mu_{S}=0$ [$\mu_{F}=-eV$]. We denote
with $\sigma\in\{\uparrow,\downarrow\}$ an electronic spin component collinear
to the polarization of $F$. Electron states in the $\sigma$-spin band [denoted
$(e,\sigma)$] at an energy $\varepsilon$ and hole states in the $-\sigma$ spin
band [denoted $(h,-\sigma)$] at an energy $-\varepsilon$ are coupled
coherently by Andreev reflections occurring at the $SF$ interface. Using the
scattering approach\cite{Blanter}, we find
\begin{align}
S(V,\omega) &  =\frac{e^{2}}{h}\sum\limits_{\substack{M,P,\\\gamma,\eta
,\alpha,\beta}}\lambda(\alpha)\lambda(\beta)\int\nolimits_{-\infty}^{+\infty
}d\varepsilon P_{M,P}^{\gamma,\eta}(\varepsilon,\varepsilon+\omega)\nonumber\\
&  \mathrm{Tr}\left[  \mathcal{A}_{F,M,P}^{\alpha,\gamma,\eta}(\varepsilon
,\varepsilon+\hbar\omega)\mathcal{A}_{F,P,M}^{\beta,\eta,\gamma}%
(\varepsilon+\hbar\omega,\varepsilon)\right]  \,.\label{Sgal2}%
\end{align}
Here, capital Latin indices correspond to the lead $F$ or $S$ and Greek
indices correspond to the electron or hole band of the space $\mathcal{E}%
_{\sigma}=\{(e,\sigma),(h,-\sigma)\}$. We use $\mathcal{A}_{F,M,P}%
^{\gamma,\alpha,\eta}(\varepsilon,\varepsilon^{\prime})=\mathbb{I}_{F,\gamma
}\delta_{F,M}\delta_{F,P}\delta_{\gamma,\alpha}\delta_{\gamma,\eta}-\left(
\mathcal{S}_{F,M}^{\gamma,\alpha}(\varepsilon)\right)  ^{\dag}\mathcal{S}%
_{F,P}^{\gamma,\eta}(\varepsilon^{\prime})$, where $\mathcal{S}_{F,M}%
^{\gamma,\alpha}(\varepsilon)$ is the scattering matrix accounting for the
transmission of quasiparticles with energy $\varepsilon$ from a state of type
$\alpha$ in lead $M$ to a state of type $\gamma$ in lead $F$. We define
$\lambda\lbrack(e,\sigma)]=-1$, $\lambda\lbrack(h,-\sigma)]=+1$ and
$P_{M,P}^{\gamma,\eta}(\varepsilon,\varepsilon^{\prime})=\left[
1-f_{M}^{\gamma}(\varepsilon)\right]  f_{P}^{\eta}(\varepsilon^{\prime})$ with
$f_{M}^{\gamma}(\varepsilon)=\left(  1+\exp[\left(  \varepsilon+\lambda
\lbrack\gamma]\mu_{M}\right)  /k_{B}T]\right)  ^{-1}$, and $T$ the
temperature. We note $e=\left|  e\right|  $ the absolute value of the electron
charge and $\mathbb{I}_{F,\gamma}$ the identity matrix in the subspace of
states of type $\gamma$ of lead $F$. In the limit $T=0$, using the symmetry of
the BdG equations (see Ref.~\onlinecite{zfQPC}, appendix A) and the unitarity
of $\mathcal{S}(\varepsilon)$, one finds
\begin{align}
S_{ex}(\omega) &  =S_{d}(\omega)+\frac{e^{2}}{h}\left(  \int\nolimits_{-\hbar
\left|  \omega\right|  }^{0}d\varepsilon\mathrm{Tr}[\widetilde{F}%
_{1}(\varepsilon,\omega,V)]\right.  \nonumber\\
&  +\left.  \int\nolimits_{-e\left|  V\right|  }^{-\hbar\left|  \omega\right|
}d\varepsilon\mathrm{Tr}[\widetilde{F}_{2}(\varepsilon,\omega,V)]\right)
\label{S1}%
\end{align}
for $0<\left|  \hbar\omega\right|  <e\left|  V\right|  $,
\begin{equation}
S_{ex}(\omega)=S_{d}(\omega)+\frac{e^{2}}{h}\int\nolimits_{-e\left|  V\right|
}^{e\left|  V\right|  -\hbar\left|  \omega\right|  }d\varepsilon
\mathrm{Tr}[\widetilde{F}_{1}(\varepsilon,\omega,V)]\label{S2}%
\end{equation}
for $eV<\left|  \hbar\omega\right|  <2e\left|  V\right|  $, and $S_{ex}%
(\omega)=S_{d}(\omega)$ for $\left|  \hbar\omega\right|  >2e\left|  V\right|
$, with
\begin{equation}
S_{d}(\omega)=\frac{2e^{2}}{h}\theta(-\omega)\int\nolimits_{-e\left|
V\right|  }^{0}d\varepsilon\operatorname{Re}\left[  \mathrm{Tr}\left[
\widetilde{F}_{3}(\varepsilon,\omega,V)\right]  \right]  \label{y2}%
\end{equation}
and $\theta(\omega)=(1+\mathrm{sign}(\omega))/2$. We define, for with
$i\in\{1,2,3\}$, $\widetilde{F}_{i}(\varepsilon,\omega,V)=\sum
\nolimits_{\sigma}F_{i,\sigma}^{\mathrm{sign}[V]\varepsilon,\mathrm{sign}%
[V]\left|  \omega\right|  }$, with
\begin{align}
F_{1,\sigma}^{\varepsilon,\omega} &  =s_{ee}^{\sigma}(\varepsilon
)s_{ee}^{\sigma\dag}(\varepsilon)s_{eh}^{\sigma}(\varepsilon+\hbar
\omega)s_{eh}^{\sigma\dag}(\varepsilon+\hbar\omega)\nonumber\\
&  -\operatorname{Re}[s_{ee}^{\sigma}(\varepsilon)s_{he}^{\sigma\dag
}(\varepsilon)s_{hh}^{\sigma}(\varepsilon+\hbar\omega)s_{eh}^{\sigma\dag
}(\varepsilon+\hbar\omega)]\label{J}%
\end{align}%
\begin{align}
F_{2,\sigma}^{\varepsilon,\omega} &  =2\operatorname{Re}[s_{ee}^{\sigma\dag
}(\varepsilon)s_{ee}^{\sigma}(\varepsilon+\hbar\omega)s_{he}^{\sigma\dag
}(\varepsilon+\hbar\omega)s_{he}^{\sigma}(\varepsilon)]\label{F}\\
&  +\sum\limits_{a\in\{e,h\}}s_{ae}^{\sigma}(\varepsilon)s_{ae}^{\sigma\dag
}(\varepsilon)\left(  \mathbb{I}_{\sigma}-s_{ae}^{\sigma}(\varepsilon
+\hbar\omega)s_{ae}^{\sigma\dag}(\varepsilon+\hbar\omega)\right)  \nonumber
\end{align}
and%
\begin{equation}
F_{3,\sigma}^{\varepsilon,\omega}=\sum\limits_{a\in\{e,h\}}\lambda_{a}%
s_{ae}^{\sigma\dag}(\varepsilon)[s_{ae}^{\sigma}(\varepsilon-\hbar
\omega)-s_{ae}^{\sigma}(\varepsilon+\hbar\omega)].\label{L}%
\end{equation}
We have used above $s_{ab}^{\sigma}(\varepsilon)=S_{F,F}^{p(a,\sigma
),p(b,\sigma)}(\varepsilon)$, with $p(e[h],\sigma)=(e[h],\pm\sigma)$,
$\lambda_{e[h]}=\mp1$, and $\mathbb{I}_{\sigma}=\mathbb{I}_{F,(e,\sigma)}$. At
last, $S$ can be obtained from $S_{ex}$ using
\begin{equation}
\frac{S(V=0,\omega)h}{2e^{2}}=\theta(-\omega)\sum\limits_{\sigma}%
\int\nolimits_{0}^{-\hbar\omega}d\varepsilon\operatorname{Re}\left[
\mathrm{Tr}[K_{\sigma}^{\omega}(\varepsilon)]\right]  \label{Vnul}%
\end{equation}
with $K_{\omega}^{\sigma}(\varepsilon)=\mathbb{I}_{\sigma}-\sum\nolimits_{a}%
\lambda_{a}s_{ae}^{\sigma\dag}(\varepsilon)s_{ae}^{\sigma}(\varepsilon
+\hbar\omega)$.\begin{figure}[ptbh]
$\qquad%
\begin{tabular}
[c]{|c|c|}\hline
$\circ$ & $-eV-\widetilde{\Delta}$\\\hline
$\vartriangleleft$ & $-\Delta-\widetilde{\Delta}$\\\hline
\end{tabular}
~
\begin{tabular}
[c]{|c|c|}\hline
$+$ & $-\Delta$\\\hline
$\boxplus$ & $-\widetilde{\Delta}$\\\hline
\end{tabular}
~
\begin{tabular}
[c]{|c|c|}\hline
$\square$ & $\widetilde{\Delta}-eV$\\\hline
$\times$ & $0$\\\hline
\end{tabular}
~
\begin{tabular}
[c]{|c|c|}\hline
$\bigstar$ & $eV-\Delta$\\\hline
$\blacksquare$ & $eV-\widetilde{\Delta}$\\\hline
\end{tabular}
~
\begin{tabular}
[c]{|c|c|}\hline
$\blacktriangleleft$ & $\Delta+\widetilde{\Delta}$\\\hline
$\bullet$ & $eV+\widetilde{\Delta}$\\\hline
\end{tabular}
$\newline \bigskip\ \ \ \ \ \ \ \ \ \newline \vspace*{0.2cm} \includegraphics
[width=0.8\linewidth]{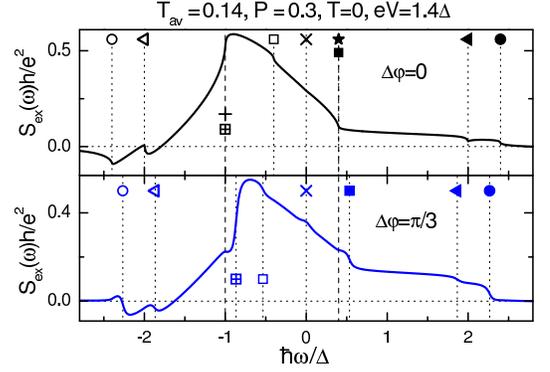}\caption{Excess noise $S_{ex}$ versus $\omega$,
for $V=1.4\Delta$, $T=0$, and different values of the SDIPS parameter
$\Delta\varphi$. The symbols correspond to frequencies defined in the above
table.}%
\label{Fig2}%
\end{figure}

In the following, we consider a single channel $SF$ QPC, which can be modeled
as a clean $SN$ contact in series with a $NF$ contact, with the length of $N$
tending to zero\cite{BeenakkerSNN}. The scattering matrix $\mathcal{P}%
^{\sigma}$ of the $NF$ contact can be expressed as $\mathcal{P}_{FF(NN)}%
^{\sigma}=r_{\sigma}\exp[i\varphi_{FF(NN)}^{\sigma}]$ and $\mathcal{P}%
_{NF(FN)}^{\sigma}=t_{\sigma}\exp[i\varphi_{NF(FN)}^{\sigma}]$, with
$r_{\sigma}^{2}+t_{\sigma}^{2}=1$ and $\varphi_{NN}^{\sigma}+\varphi
_{FF}^{\sigma}=\varphi_{NF}^{\sigma}+\varphi_{FN}^{\sigma}+\pi~[2\pi]$. In
general, the scattering phases $\varphi_{ij}^{\sigma}$, with $(i,j)\in
\{N,F\}^{2}$, are spin-dependent because the QPC displays a spin-dependent
scattering potential. This so-called Spin-Dependence of Interfacial Phase
Shifts (SDIPS) modifies qualitatively the behavior of the QPC. A step
approximation for the gap $\Delta$ of $S$ leads to an $SN$ Andreev reflection
amplitude $\gamma=(\varepsilon-\mathrm{i}\sqrt{\Delta^{2}-\varepsilon^{2}%
})/\Delta$ for $\left|  \varepsilon\right|  <\Delta$ and $\gamma
=(\varepsilon-\mathrm{sgn}(\varepsilon)\sqrt{\varepsilon^{2}-\Delta^{2}%
})/\Delta$ for $\left|  \varepsilon\right|  >\Delta$\cite{BTK}. The resulting
scattering matrix $\mathcal{S}(\varepsilon)$ can be found in Ref.
\onlinecite{zfQPC}, Appendix B. Importantly, the elements of $\mathcal{S}%
(\varepsilon)$ involve the denominator $D_{\sigma}(\varepsilon)=1-\gamma
^{2}r_{\sigma}r_{-\sigma}e^{i(\varphi_{NN}^{\sigma}-\varphi_{NN}^{-\sigma})}$,
which accounts for iterative reflection processes between the $SN$ and $NF$
interfaces (Andreev bound states). A quasiparticle can interfere with itself
after two back and forth travels between $S$ and $F$, one as an electron
$(e,\sigma)$ and one as a hole $(h,-\sigma)$, which leads to a dependence of
$D_{\sigma}$ on the SDIPS parameter $\Delta\varphi=\varphi_{NN}^{\uparrow
}-\varphi_{NN}^{\downarrow}$. We have checked analytically that $S(V,\omega)$
depends on the phases of $\mathcal{P}^{\sigma}$ through $\Delta\varphi$ only,
for all values of $\omega$, $V$ and $T$. For simplicity, we will consider
below that $t_{\sigma}$ and $\Delta\varphi$ are independent of $\varepsilon$
and $V$, so that the energy dependence of $\mathcal{S}(\varepsilon)$ stems
only from $\gamma$. We will mainly focus on the regime of weak transparency
$T_{av}=(t_{\uparrow}^{2}+t_{\downarrow}^{2})/2\ll1$, with a finite
polarization $P=(t_{\uparrow}^{2}-t_{\downarrow}^{2})/(t_{\uparrow}%
^{2}+t_{\downarrow}^{2})$. We will use $T=0$ except in Fig.~\ref{Fig3}, right panel.

\begin{figure}[ptbh]
\vspace*{0.2cm}\includegraphics[width=0.85\linewidth]{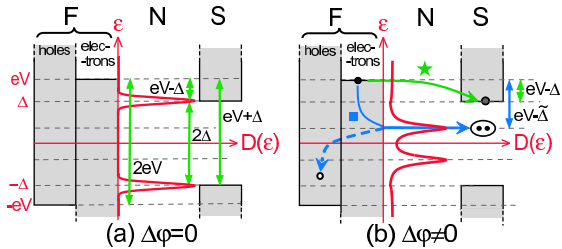}\caption{Energy
diagram of the $SF$ $QPC$, for $\Delta\varphi\neq0$, $eV>\Delta$,
$\Delta\varphi=0$ (panel a) and $\Delta\varphi\neq0$ (panel b). The various
columns represent the (hole) electron band in $F$, filled (down) up to the
energy $(-)eV$, and the density of states in $S$, which presents a gap for
$\varepsilon\in\lbrack-\Delta,\Delta]$. The resonance function $D(\varepsilon
)=\left|  D_{\uparrow}(\varepsilon)\right|  ^{-2}+\left|  D_{\downarrow
}(\varepsilon)\right|  ^{-2}$ is shown with red full lines. In panel (a), the
different frequencies $\omega$ at which $S_{ex}$ is singular are represented
with green arrows. In panel (b), we show with blue and green arrows transfer
processes contributing to the $\bigstar$ and $\blacksquare$ features of
Fig.\ref{Fig2}.}%
\label{Fig5}%
\end{figure}

We now study the frequency-parity of $S_{ex}$. For $0<e\left|  V\right|
,\left|  \hbar\omega\right|  \ll\Delta$, the function $S_{d}(\omega)$ can be
neglected from Eqs.~(\ref{S1}-\ref{S2}), so that one finds $S_{ex}%
(\omega)=s_{0}(\left|  V\right|  -\hbar\left|  \omega/2e\right|
)\theta(2e\left|  V\right|  -\hbar\left|  \omega\right|  )$, with
$s_{0}=4(e^{3}/h)\sum\nolimits_{\sigma}\mathrm{Tr}[s_{he}^{\sigma\dag
}(0)s_{he}^{\sigma}(0)\left(  1-s_{he}^{\sigma\dag}(0)s_{he}^{\sigma
}(0)\right)  ]$ the zero-frequency differential noise. In this regime,
$S_{ex}$ is thus is an even function of $\omega$, determined by zero-frequency
properties only. The behavior of $S_{ex}$ is much richer for $eV>\Delta$.
Fig.~\ref{Fig2} shows $S_{ex}$ versus $\omega$, for $eV=1.4\Delta$ and
different values of $\Delta\varphi$. We find that $S_{ex}$ is not an even
function of $\omega$ anymore, because Andreev reflections make $\mathcal{S}%
(\varepsilon)$ strongly energy-dependent [see Eq.~(\ref{L})]\cite{note2}.
Hence, it can be useful to measure $S_{ex}$ for both signs of $\omega$, to
gain additional information on the QPC. Interestingly, $S_{ex}(\omega)\neq$
$S_{ex}(-\omega)$has already been observed in a Josephson
junction\cite{Billangeon}. We now focus on the sign of $S_{ex}$. For
$\omega>0$ and $T=0$, one always finds $S_{ex}\geq0$, because emission
processes are forbidden for $V=0$ [see Eq.~(\ref{Vnul})], so that $S_{ex}=S$
at positive frequencies. However, nothing forbids to have $S_{ex}<0$ for
$\omega<0$, as illustrated by Fig.~\ref{Fig2}. Note, that $S_{ex}<0$ has
already been predicted for a non-interacting conductor with energy-dependent
transmission probabilities \cite{lesovik} and an interacting quantum wire
\cite{Dolcini}. However, in experimental works such a behavior has not been
reported yet.

We now study the frequency dependence of $S_{ex}$. For $\Delta\varphi=0$ (top
panel of Fig.~\ref{Fig2}), $S_{ex}$ shows various singularities, which can be
understood from Eq.~(\ref{Sgal2}). The current noise at frequency $\omega$
involves a coupling between quasiparticles states with energy $\varepsilon
_{1}=\varepsilon$ and $\varepsilon_{2}=\varepsilon+\hbar\omega$. The
occupation of states in $F$ is discontinuous at $\varepsilon=\pm eV$, and the
density of states in $S$ is singular at $\varepsilon=\pm\Delta$. This allows
singularities of $S_{ex}$ at frequencies $\omega=(\varepsilon_{1}%
-\varepsilon_{2})/\hbar$ with $(\varepsilon_{1},\varepsilon_{2})\in
\{eV,-eV,\Delta,-\Delta\}^{2}$ i.e. $\hbar\left|  \omega\right|
\in\{0,\left|  \Delta\pm eV\right|  ,2\Delta,2e\left|  V\right|  \}$, as
already found by Ref.~\onlinecite{Torres} for a $S/N$ QPC with $\omega
>0$\cite{cancel}. These different frequencies can be represented on the energy
diagram of the circuit (see Fig. \ref{Fig5},a for the case $eV>\Delta$). In
Fig.~\ref{Fig2}, we find that $S_{ex}$ displays an extra noise singularity at
$\hbar\omega=-\Delta$, because it involves $S(V=0,\omega)$. For $\Delta
\varphi\neq0$, the denominator $D_{\sigma}(\varepsilon)$ is resonant at subgap
energies $\varepsilon=\sigma\widetilde{\Delta}$ with $\widetilde{\Delta
}=\Delta\cos(\Delta\varphi/2)$. This reveals that subgap resonances occur in
the N layer\cite{zfQPC}. This effect modifies qualitatively the behavior of
the QPC: in Fig. \ref{Fig2}, bottom panel, noise steps appear at $\hbar\left|
\omega\right|  \in\{\left|  \widetilde{\Delta}\pm eV\right|  ,\Delta
+\widetilde{\Delta}\}$ and $\hbar\omega=-\widetilde{\Delta}$, marked with the
symbols $\circ$, $\vartriangleleft$, $\boxplus$, $\square$, $\blacksquare$,
$\blacktriangleleft$, and $\bullet$. These steps are due to particle transfer
processes which involve the subgap resonance in the N layer. For instance, the
step occurring at $\omega=eV-\widetilde{\Delta}$, marked with $\blacksquare$
in Fig. \ref{Fig2}, corresponds e.g. to the Andreev reflection of electrons
with energy $eV$ of lead F, via the Andreev subgap resonance at energy
$\widetilde{\Delta}$ (see Figure \ref{Fig5},b, blue arrow). Importantly, for
$\Delta\varphi=0$, the SDIPS-induced noise steps are absorbed into the noise
singularities of the $\Delta\varphi=0$ case which are thus reinforced, because
$\left|  D_{\sigma}(\varepsilon)\right|  ^{2}$ is resonant at $\varepsilon
=\pm\Delta$. When $\Delta\varphi\neq0$, the singularities which existed for
$\Delta\varphi=0$ are either smoothed, like the singularity at $\hbar
\omega=eV-\Delta$ (marked with $\bigstar$) and the singularity at $\hbar
\omega=-\Delta$ (marked with $+$), or even vanish. Measuring the frequencies
of the steps appearing in $S_{ex}$ can give a direct access to the SDIPS
parameter $\Delta\varphi$, since these frequencies are merely linear
combinations of $\widetilde{\Delta}$, $V$, and $\Delta$, independently of any
other parameter. Moreover, $V$ does not affect the positions of all the
singularities and steps of $S_{ex}$ in the same manner (see table in
Fig.~\ref{Fig2}), which can help to identify these features in an experiment.
In practice, the frequency range studied in Fig.~\ref{Fig2} should be
accessible with on-chip detectors, which presently allow to reach $\omega
\sim250~$\textrm{GHz}\cite{Onac,freq}.\begin{figure}[ptbh]
\includegraphics[width=0.8\linewidth]{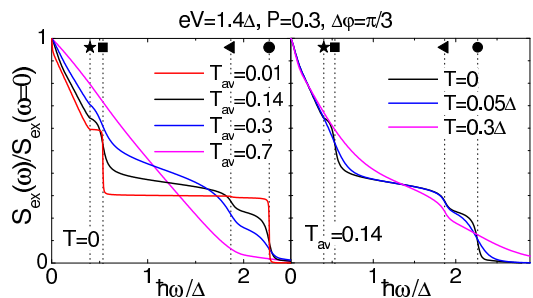}\caption{Excess noise $S_{ex}$
versus $\omega$. The left and right panels show the effect of an increase of
the spin-averaged transmission probability $T_{av}$ and of the temperature
$T$, respectively. The symbols correspond to frequencies defined in Fig.
\ref{Fig2}.}%
\label{Fig3}%
\end{figure}

The left and right panels of Fig.~\ref{Fig3} show how $S_{ex}$ depends on the
spin-averaged transmission probability $T_{av}$ of the QPC and on the
temperature $T$, respectively. For brevity, we present results for $\omega>0$
only. The singularities and steps displayed by $S_{ex}$ remain visible for
experimentally accessible temperatures (see right panel, blue curve, for
$k_{B}T=0.05\Delta$). Remarkably, $T$ and $T_{av}$ do not influence the
visibility of all the steps and singularities of $S_{ex}$ in the same way.
Those occurring for $\hbar\omega\in\{eV-\Delta,eV\mp\widetilde{\Delta}\}$,
marked with $\bigstar$, $\blacksquare$ and $\bullet$, are very sharp when
$T_{av}$ and $T$ are small (see left panel, red curve). The visibility of
these features decrease when $T_{av}$ or $T$ increase, due to a smoothing of
$\left|  D_{\sigma}(\varepsilon)\right|  ^{2}$ or of the Fermi factors,
respectively. The step occurring at $\hbar\omega=\Delta+\widetilde{\Delta}$,
marked with $\blacktriangleleft$, has a significantly different behavior.
First, it is smoothed by an increase in $T$, but much more slowly than the
other singularities. This is due to the fact that the dependence of $S$ on
$\Delta$ and $\widetilde{\Delta}$ occurs through the scattering matrix of the
QPC, which does not depend on $T$, whereas $V$ occurs through the Fermi
factors. Second, this step vanishes for $T_{av}$ large but also for $T_{av}$
small. From a detailled analysis of Eq.~(\ref{J}), this last behavior can be
traced to the fact that $s_{ee}^{\sigma}(\varepsilon)$ is almost constant
around $\varepsilon=\Delta$ at low transparencies.\begin{figure}[ptbh]
\includegraphics[width=0.85\linewidth]{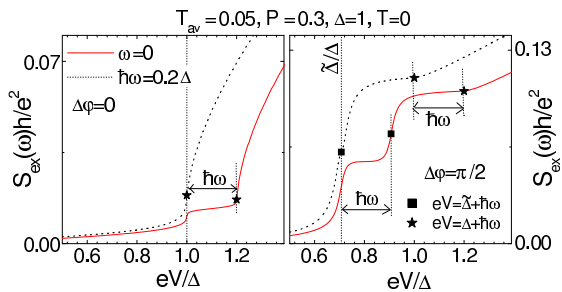}\caption{{}Excess noise
$S_{ex}$ as a function of the bias voltage $V$, for $\omega=0$ $[\omega\neq0]$
(dotted [full] lines). We have used no SDIPS on the left panel and a finite
SDIPS on the right panel.}%
\label{Fig4}%
\end{figure}

Finally, we study the dependence of $S_{ex}$ on the bias voltage $V$, which
can be measured e.g. using cryogenic low noise amplifiers \cite{Christian}.
The left and right panels of Fig.~\ref{Fig4} show curves calculated for a
weakly transparent contact, without and with a SDIPS respectively. For
simplicity, we focus on the range $eV<2\Delta$. For $\Delta\varphi=0$ and
$\omega=0$, $S_{ex}$ presents a single singularity at $V=\Delta/e$. This
singularity follows straightforwardly from Eq.~(\ref{Sgal2}), but it has not
appeared in Fig.~\ref{Fig2} because its position is independent of $\omega$.
For $\Delta\varphi=0$ and $\omega\neq0$, a second singularity appears, at
$V=(\Delta+\hbar\omega)/e$. This feature corresponds to the step at
$\hbar\omega=eV-\Delta$ in Fig.~\ref{Fig2}. When $\Delta\varphi\neq0$, a
single singularity occurs, at $V=(\Delta+\hbar\omega)/e$. It is softer than in
the case $\Delta\varphi=0$, similarly to what happens for the singularity at
$\hbar\omega=eV-\Delta$ in Fig.~\ref{Fig2}. Furthermore, two noise step
appears at $V=\widetilde{\Delta}/e$ and $V=(\widetilde{\Delta}+\hbar\omega
)/e$. Although cryogenic low noise amplifiers are presently limited to
$8~$\textrm{GHz}\cite{Christian}, they should allow to observe the features
described in this pararaph since they can occur at frequencies much lower than
$\Delta$ \cite{freq}. The SDIPS already modifies qualitatively the noise curve
for $\omega=0$, but we think that a measurement at $\omega\neq0$ will be
crucial in practice. It will allow to confirm the nature of the singularities
and steps occurring in $S_{ex}$, and it will also help to calibrate the
observed $\widetilde{\Delta}$ and $\Delta$, thanks to the scale $\hbar\omega$
which appears naturally in the noise curves (see Fig.~\ref{Fig4}, horizontal arrows).

So far, experiments on $S/F$ point contacts have been performed in the limit
of a large number of channels. However, we believe that the few channels
regime may be realized, using e.g. a Scanning Tunneling Microscope (STM).
Indeed, STMs allow to reach the few channels regime (atomic contact)
\cite{Rodrigo}, and they are particularly suitable for making hybrid contacts.
Moreover, current noise has already been measured in a STM\cite{Birk}. Another
possibility could be to contact a carbon nanotube to a $F$\cite{SST} and a
$S$\cite{Kasumov}. Interestingly, a multichannel ballistic $S/F$ contact has
been considered in Ref.~\onlinecite
{multichannel}, using a rectangular potential barrier model. Although the
transverse momentum of particles has been taken into account, the SDIPS is
found to be approximately independent of the channel index when $F$ is
strongly polarized\cite{zfQPC}. We think that the SDIPS-induced noise steps
are bound to persist in these conditions. More generally, in the multichannel
case, the occurence of the SDIPS-induced noise steps will depend on the
detailed structure of the contact.

In summary, we have calculated the finite-frequency non-symmetrized current
noise of a superconductor/ferromagnet quantum point contact (QPC). For low
transparencies, this quantity is qualitatively modified by the Spin-Dependence
of Interfacial Phase Shifts, which induces characteristic noise steps at
various frequencies and voltages. Finite-frequency noise can thus be an
interesting tool to characterize the scattering properties of a $SF$ QPC.

We acknowledge financial support by the Swiss National Science Foundation, the
German Research Foundation (DFG) through SFB 767 and the Landesstiftung
Baden-W\"{u}rttemberg. We thank C. Glattli and T. Kontos for discussions.


\begin{thebibliography}{99}
\bibitem{note2}For $N/N$ QPCs, $S_{ex}(\omega)=S_{ex}(-\omega)$ is often found
(see e.g. Refs. \onlinecite{Deblock,Martin}), due to the hypothesis of energy
independent scattering [see Eq.~(\ref{L})].
\bibitem{cancel}The singularity at $\hbar\omega=2eV$ always occurs because it
is due to the cancellation of $S_{ex}$ at $\omega$ large. However, this
feature is outside the range of Fig.~\ref{Fig2}.
\bibitem{freq}If $S$ is e.g. an \textrm{Al} electrode, one has $\Delta\sim
46~$\textrm{GHz}.
\end{thebibliography}
\end{document}